\documentclass{article}
\usepackage[utf8]{inputenc}
\usepackage{authblk}
\usepackage{setspace}
\usepackage[margin=1.25in]{geometry}
\usepackage{graphicx}
\graphicspath{ {./figures/} }
\usepackage{subcaption}
\usepackage{amsmath}
\usepackage{lineno}
\usepackage{longtable}
\linenumbers

\usepackage[style=nejm, 
citestyle=numeric-comp,
sorting=none]{biblatex}
\title{Spectral Umixing Comparison with Sparse, Iterative and Mixed Integer Programming Models}

\author[1$\dag$]{Jade F. Preston}
\author[1*$\dag$]{William F. Basener}

\affil[1]{School of Data Science, University of Virginia, Charlottesville, VA.}
\affil[*]{Address correspondence to: wb8by@virginia.edu}

\date{}

\begin{document}
\nolinenumbers

\maketitle

\begin{abstract}
Hyperspectral unmixing is the analytical process of determining the pure materials and estimating the proportions of such materials composed within an observed mixed pixel spectrum. We can unmix mixed pixel spectra using linear and nonlinear mixture models. Ordinary least squares (OLS) regression serves as the foundation for many linear mixture models employed in Hyperspectral Image analysis. Though variations of OLS are implemented, studies rarely address the underlying assumptions that affect results. This paper provides an in depth discussion on the assumptions inherently endorsed by the application of OLS regression. We also examine variations of OLS models stemming from highly effective approaches in spectral unmixing -- sparse regression, iterative feature search strategies and Mathematical programming. These variations are compared to a novel unmixing approach called HySUDeB. We evaluated each approach's performance by computing the average error and precision of each model. Additionally, we provide a taxonomy of the molecular structure of each mineral to derive further understanding into the detection of the target materials.
\end{abstract}


\section{Introduction}
Hyperspectral images capture the electromagnetic radiation reflected from surfaces and record the reflectance across hundreds or thousands of contiguous spectral bands \cite{manolakis2003hyperspectral}. Researchers refer to this mapping of material reflectance across spectral bands as spectral signatures or spectra. Pixels within the images frequently contain a mixture of different materials -- ultimately modifying the amount of light photons returning to the hyperspectral sensor \cite{somers2011endmember}. Researchers aim to identify and quantify these materials through a process called spectral unmixing. Material mixtures can be categorized and modeled as linear or nonlinear depending on compositional complexity \cite{keshava2002spectral}.  

The linear mixture model (LMM) -- depicted by equation \ref{eq:LMM_equation} -- is often interpreted as a regression problem \cite{keshava2002spectral,borsoi2021spectral}. The observed mixed pixel spectrum (singular form of spectra) is represented by the dependent variable, $\mathbf{y}$, and can be further described as a vector of the pixel's recorded reflectance at waveband (\emph{i} to \emph{N}). We, therefore, model the observed spectrum by inferring the pure materials or endmembers from an image or constituent spectra from a  spectral library. In this study, we use an extensive spectral library for unmixing rather relying on ``pure" image pixels. The matrix of constituent spectra is represented by $\mathbf{S}$. Neither the image endmembers nor the abundances are  known \emph{a priori}. However, if the spectral library is small, we can invert the library to estimate these unknown abundances, represented by the vector $\mathbf{a}$ \cite{dismuke2006ordinary}. Inversion becomes impossible when the spectral library is large --- an increasingly common scenario --- particularly when there are more library spectra than spectral bands. Equation \ref{eq:LMM_solution} displays the unmixing solution estimating the unknown, fractional proportion of materials. All models inherently incorporate some level of error. Thus, the unmixing solution is an estimate indicated by the lack of error vector, $\mathbf{E}$, shown in equation \ref{eq:LMM_equation}. In least squares models, the error is referred to as the residual or in this case the distance between the inferred reflectance and the true pixel reflectance at wavelength \emph{i}.

\begin{equation}
\label{eq:LMM_equation}
    \mathbf{y} = \mathbf{Sa} + \mathbf{E}
\end{equation}

\begin{equation}
\label{eq:LMM_solution}
    \mathbf{\hat{a}} = (\mathbf{S^TS})^{-1}\mathbf{S^Ty}
\end{equation}

The basic form of linear regression is also referred to as OLS \cite{fox2015applied,burton2021ols}. OLS minimizes modeling error by reducing the sum of the squared differences between the observed and inferred reflectance values. There are alternative regression approaches (i.e. generalized or weighted least squares), but this study focuses on OLS because it serves as the baseline for most unmixing techniques \cite{dismuke2006ordinary}. A common metric for assessing model error is mean squared error (MSE), which is calculated by normalizing the residual sum of squares by the number of data points. Equation \ref{eq:MSE_equation} presents the formula for MSE. The OLS objective is to minimize the MSE by identifying the set of coefficients that maximize the similarity between the response and the regressor values \cite{Wang2009Mean}. 

\begin{equation}
\label{eq:MSE_equation}
    \text{MSE} = \frac{1}{M}\sum_{i=1}^M(\mathbf{y_{i}} - \mathbf{(Sa)_{i}})^2 
\end{equation}

 The scope of this study is primary material identification via four unmixing methods. We aim to identify specific substances embedded in a mixed pixel regardless of pixel composition. In this paper, we will examine the assumptions underlying OLS and their alignment, or lack thereof, with the LMM. For instance, the OLS mixture model fails to provide practical abundance estimates when the spectral library is large. 
 
 This paper also provides a comparison focusing on four highly effective unmixing techniques: sparse regression, depth-first search (DFS) feature selection, mixed integer nonlinear programming (MINLP), and band decorrelation. In many past papers, the implementation of certain feature selection techniques --- i.e as step-wise regression --- have been critiqued. It is our additional aim to discuss the attributes of iterative feature selection that are useful in spectral unmixing. 
 
Dimension reduction and decorrelation of library spectra are common practice in spectral unmixing. To our knowledge, no other study has pre-processed the data by whitening the spectral bands as part the unmixing methodology. This paper presents a novel approach called Hyperspectral unmixing with decorrelated bands (HySUDeB). Each of the techniques discussed in this paper respond to one or more of the inadequacies realized through OLS unmixing. We compare the three previously mentioned unmixing techniques to HySUDeB and investigate their performance trade offs. This investigation includes a description of the physical-chemical structure of the materials aiding in accurate detection.

\subsection{Ordinary Least Square Assumptions}
Researchers, in the spectral analysis field, often discuss the assumptions supporting the LMM but the assumptions behind OLS are rarely explained. The outline below highlights primary OLS assumptions and discusses their validity in the context of spectral unmixing \cite{dismuke2006ordinary, poole1971assumptions,tranmer2008multiple}: 

\begin{enumerate}
    \item Linearity in Relationships --- OLS assumes the dependent variable is a linear combination of the predictor variables and the residuals. Specifically, linearity is exhibited through the additive nature between parameters. This relationship is modeled as a linear function (shown by Equation \ref{eq:LMM_equation}) in which both the contribution of the input variables and the error term are considered additive.

    This linear assumption holds primarily for hyperspectral unmixing scenarios with macroscopic mixtures. Endmembers in these pixels occupy distinct or adjacent regions thus, we assume light photons interact with only one material before returning to the sensor. 
    
    However, this assumption fails when light photons interact with more than one surface before returning to the sensor. Scenarios where this assumption fails are when light interactions with material mixtures are more complicated (e.g. scenes with high vegetation or mineral deposits) \cite{keshava2002spectral,dobigeon2013nonlinear}.
    
    \item Independent Identically Distributed (IID) error --- The residuals from OLS are assumed to satisfy three key properties: independence, homoscedasticity, and normality with mean centered at zero. Independence implies that there is no correlation between the residuals. This independence assumption means that the error of one observation (i.e. band 1 residual) will not be correlated with the error of another observation (i.e. band 2 residual). Homoscedasticity means the variance or spread of the residuals is constant across the input variables. When we say the residuals are normally distributed with mean equaling zero, it means the residuals follow a bell-shaped curve when plotted as a histogram, and their expected value is zero. Because the expected value of the error is zero, this implies: 1. OLS is an unbiased estimator and 2. there is no correlation between the residuals and the input variables. Therefore, we assume error is not computed in any one direction. 

   In the context of the LMM, a multivariate normal distribution is assumed for the residuals with the mean centered at the observed pixel spectrum. This approximation is considered generally reasonable, but errors derived from the LMM do not always adhere to the properties of independence and homoscedasticity. Continuous and adjacent spectral bands often exhibit correlations, especially between bands in close proximity \cite{chang2013hyperspectral,chang2013progressive}. Factors such as correlated bands, spectral variability, and spatio-temporal effects can disrupt homoscedasticity. As a result, residuals may exhibit fluctuating variances or in certain cases follow entirely different distributions \cite{bioucas2013hyperspectral}.

    \item Noncollinear Regressor Variables --- OLS requires the regressor variables exhibit noncollinearity, meaning there is no linear dependency between the input vectors. When the input vectors are linearly independent, a unique solution can exist which is considered vital to obtain the closed form solution presented in Equation \ref{eq:LMM_solution}.
   
   Noncollinearity is possible when there are more spectral bands than library spectra. However, when there are more library spectra than spectral bands a unique solution of abundances cannot be found. If the number of library spectra exceeds the number of bands, inversion is impossible resulting in infinite solutions. This means that any number of abundance combinations can reproduce the observed spectrum (y).

    \item Correctly established models --- The input variables are considered fixed implying that the information available to describe the response variable to known \emph{a priori}. We assume that the pertinent variables are included in the model or otherwise excluded. Additionally, the implementation of OLS regression assumes the inherent OLS assumptions are valid.

    Unmixing requires either the observed image or a spectral library. Both of these approaches can challenge the correctly established modeling assumption as well as the preceding assumptions. When unmixing a pixel using the image, we must assume that pure pixels of individual materials are present and can be mapped to the mixed pixel, or that there exists a pixel with similar spectral properties, indicating identical materials and or mixtures. Alternatively, unmixing with a spectral library inherently assumes that the library contains materials that match those in the pixel we aim to unmix. Any one of the assumptions discussed can fail causing large estimated error or erroneous modeling conclusions.
    
\end{enumerate}

\subsection{Practical LMM Constraints}
The unconstrained OLS model provides unrealistic abundance results and requires the implementation of one or more practical constraints. The two primary constraints enforced are the nonnegative abundance and abundance sum to one constraints \cite{keshava2000algorithm, heinz2001fully}. Negative abundance observed from the unconstrained model stem from a combination of over-fitting of the observed variable and the minimal presences of a particular substance \cite{winter2003examining}. The nonnegative abundance constraint -- commonly referred to as nonnegative least squares (NNLS) -- ensures any variables included in the model have coefficients greater than or equal to zero. Equation \ref{NN-constraint} displays the nonnegative abundance constraint regularly imposed on the OLS model. $N$ represents the explanatory variables. In the paper by Keshava et al., the authors discuss the implementation of the sum-to-one abundance constraint. This constraint forces the estimated coefficients of the explanatory variables to sum to one, adhering to the practical assertion that the materials in the model make up a fractional proportion of the area in the pixel \cite{keshava2002spectral}. Within this paper, we relax this constraint to comply with an alternative assumption: materials in a mixed pixel may appear brighter in certain scenarios causing the abundances to sum greater than one. Thus, the only practical constraint we enforce in this study is nonnegative abundances.

\begin{equation}
\label{NN-constraint}    
a_i \geq 0  \hspace{1cm}\forall\hspace{.2cm}1\leq i \leq N
\end{equation}

\subsection{Alternative Unmixing Techniques}
\subsubsection{Sparse Regression}

Sparse unmixing is a technique in spectral unmixing that addresses some of the misalignments found with OLS assumptions. High  correlation between library spectra is a significant challenge in spectral unmixing. Large spectral libraries are essential to capture spectral variability in intimate mixtures, but their use can lead to overfitting and an ill-posed inversion problem, thereby making regularization necessary. Sparse unmixing addresses these issues by selecting the optimal subset of input variables from the library to describe the observed spectrum \cite{borsoi2021spectral}. This approach introduces regularization to the OLS model through penalties on the parameters.

There are many studies that have achieved unmixing success using sparse unmixing approaches. The common penalties imposed are the $L_1$ norm (also known as LASSO regression), $L_2$ norm (also known as RIDGE regression) or a combination of the $L_1$ and $L_2$ norm  (their combination is known as ElasticNet). LASSO regression uses Manhattan distancing to shrink coefficients to zero penalizing the size of the model. Subsequently, this $L_1$ penalty adheres to the nonnegative abundance constraint. Equation \ref{eq:LASSO_solution} displays the modified version of equation \ref{eq:MSE_equation} which includes the $L_1$ penalty on the model coefficients. The set of coefficients, $\hat{a}$, minimizes the objective function subject to the $L_1$ penalty norm.

\begin{equation}
\label{eq:LASSO_solution}
     \mathbf{\hat{a}} = \text{ arg min}_{a}\left( \frac{1}{M}\|\mathbf{y - Sa} \|_2^2 + \lambda \| \mathbf{a} \|_1 \right)
\end{equation}

\subsubsection{Iterative Approaches}
An alternative to sparse approaches are iterative approaches i.e. depth- and breadth-first search strategies. Depth- and breadth-first search strategies are often incorporated in graph algorithms or decision trees \cite{kozen1992depth, bundy1984breadth}, but their underlying principles have been adapted to a wide variety of use cases i.e. step-wise regression. Both strategies iteratively identify a subset of features that best fit the data \cite{dash1997feature}. Step-wise regression is primarily implemented for model selection with a secondary aim of parameter estimation. In the context of unmixing, step-wise regression responds to the goal of sparse unmixing which is to identify the subset of explanatory variables from a far larger set of materials \cite{ruengvirayudh2016comparing,olusegun2015identifying}. 

Many statisticians caution against the implementation of step-wise regression due to issues with multi-colinearity. When explanatory variables are highly correlated, step-wise regression models struggle with deciphering variable influence on the response variable. R.D Routledge says that in certain scenarios of high correlation, the combination of explanatory variables can show a significant impact on the response variable, while individually the variables are insignificant \cite{routledge1990stepwise}. Additionally, in another paper by Michael Lewis-Beck, the author explains an alternative scenario where three variables are considered significant individually, but as they are iteratively included in the model, their significance lowers. He claims the iterative process fails to depict the relationship between the variables and ultimately their level of significance \cite{lewis1978stepwise}.

The studies discussing the limitations of step-wise regression often exhibit two characteristics that are less relevant to spectral unmixing. Many of the correlated variables in spectral libraries are the variations of the same substance rather than two highly correlated but different variables. Additionally, OLS variables can have a negative impact on the response variable. Practicality prohibits the inference of negative abundance coefficients in the LMM. Considering these differences, stepwise regression’s disadvantages become less pronounced in the spectral unmixing context. The primary aim of variable selection aligns with stepwise regression’s design, making it well-suited to identifying the most statistically significant materials influencing pixel reflectance. Winter et al. explains that the iterative process of step-wise regression coupled with nonnegative abundance enforcement lowers modeling error \cite{winter2003examining}. In another paper, Gault et al. compares NNLS, sparse and step-wise approaches. The authors explain that step-wise regression can out perform non-iterative approaches especially when detection is the main goal \cite{gault2016comparing}.

With these considerations in mind, we implement an iterative search strategy, focusing specifically on DFS feature selection. This approach begins with an empty set, iteratively identifies the most significant explanatory variables and incorporates them into the model. The significance of each variable is evaluated using the F-statistic, a ratio that compares explained versus unexplained variance. The F-statistic is then used to calculate the probability (p-value) of obtaining the observed F-statistic by chance. A low p-value suggests that the observed F-statistic is unlikely to have occurred randomly. When the p-value falls below a predetermined threshold, the variable is considered to contribute significantly to the variance observed in the response variable and subsequently included in the model \cite{olusegun2015identifying, routledge1990stepwise,lewis1978stepwise}.

\subsubsection{MINLP}
Despite its potential for higher accuracy \cite{mhenni2018spectral}, few researchers employ mathematical programming for spectral unmixing. Among the studies that have utilized mathematical programming, there is a noticeable lack of consistent terminology. Many papers describe techniques that originate from the broader field of mathematical programming, such as linear programming \cite{bateson2000endmember, marinoni2015novel}, integer programming \cite{joker2008exact,liu2018convex}, mixed integer programming (MIP), MINLP \cite{schmidt2021spectral, schmidt2023mineral} and other variations of nonlinear programming \cite{dache2023exact,ermon2015pattern, bourguignon2015exact}. While these techniques are related, each employs slightly different approaches. Mathematical programming is a field of study that analyzes optimal decision making subject to a set of limitations. Solutions are derived by optimizing an objective -- a minimization or maximization function -- composed of decision variables. These variables are iteratively adjusted within the defined limits to determine the best solution. The limits imposed are called constraints and represent practical requirements or boundaries. MIP integrates concepts from both linear and integer programming because the decision variables are integer and continuous. MINLP is an optimization method originating from MIP, but incorporates nonlinear relationships \cite{smith2008tutorial}.

Bateson et al. completed a study in which they developed endmember bundles using principal component analysis and simulated annealing to determine the dimension and vertices of the data simplex. Each bundle is a collection of spectra for a single material; the author incorporated grass, tree and soil bundles. Once the bundles were created, Bateson et al. unmixed pixels using linear programming \cite{bateson2000endmember}. In another study, Bourguignon et al. compare nine variations of sparse mixture models, showing the equivalence of more traditional mixture models to MIPs. The authors concluded that with an accepted increase in computational demand, mixed integer formulations can reduce the error of abundance estimation and provide exact solutions \cite{bourguignon2015exact}. Mhenni et al. echo these sentiments in their paper where they conduct unmixing using a mathematical program incorpating sparse and exclusive group constraints \cite{mhenni2018spectral}.

\section{Materials and Methods}

\subsection{Experimental Design}
Our observed dataset was derived from a hyperspectral image captured using the Airborne Visible/Infrared Imaging Spectrometer (AVIRIS). AVIRIS is one of the first airborne hyperspectral imagers developed in the 1980s by NASA's Jet Propulsion Laboratory \cite{qian2021hyperspectral}. It currently collects data across 224 contiguous spectral bands, covering wavelengths from 0.38 to 2.5 micrometers. The image used in this study consists of 350x400 pixels, capturing a section of Cuprite Hills, Nevada \cite{swayze2014mapping}. Cuprite Hills is a well-known mineralogical site and a prime testing ground for studying natural intimate mixtures. Additionally, we utilized the spectral library from the U.S. Geological Survey (USGS) to infer materials embedded in the observed Cuprite image pixels. Our spectral library was comprised of 481 mineral spectra. Many of the spectra from the library are a variation of the same substance to aid material detection within the intimate mixtures found at Cuprite Hills \cite{clark2007usgs}.

\begin{figure}[ht]
   \begin{center}
\includegraphics[width=\textwidth]{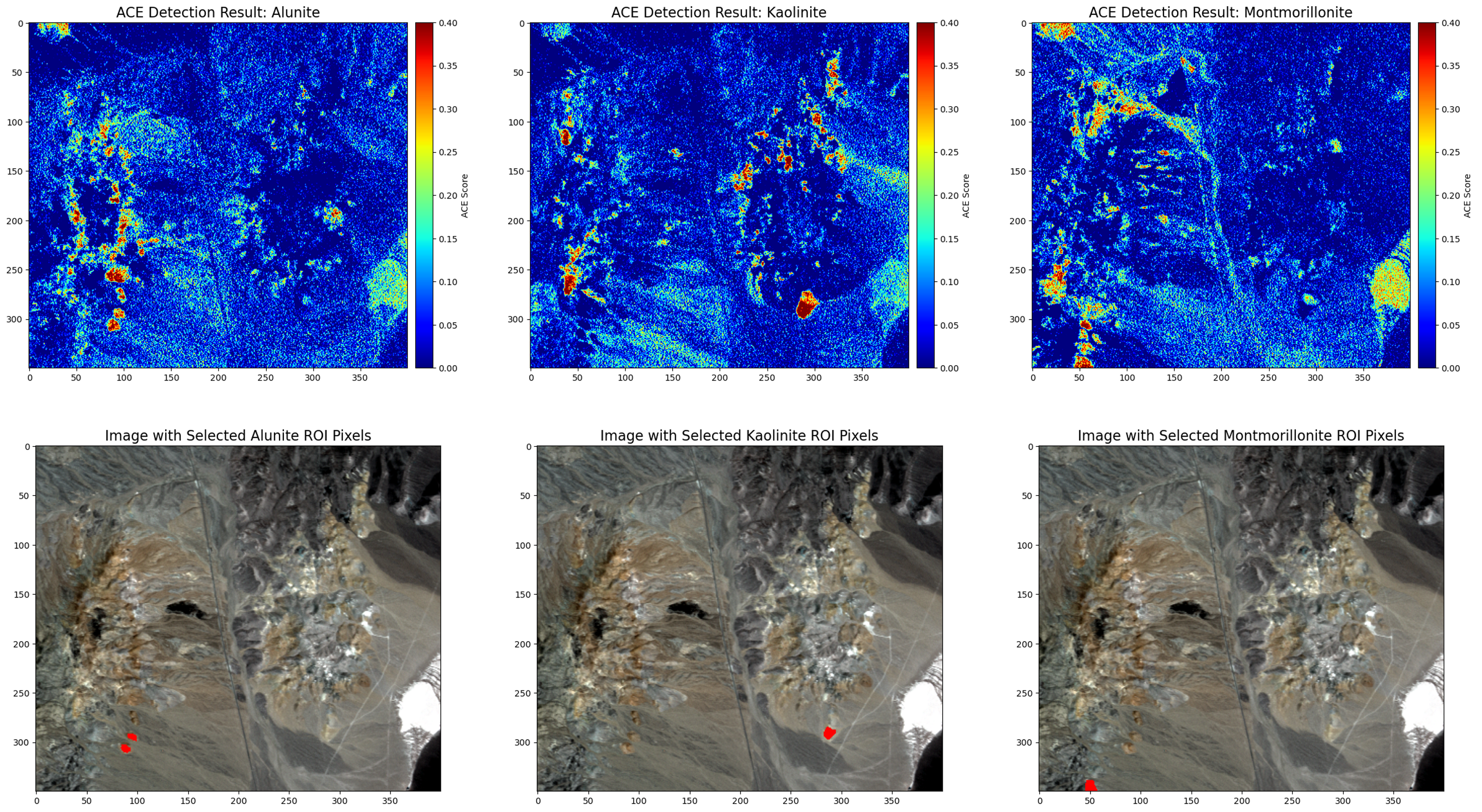}
	\end{center}
 \caption{Display of the selected ROIs for three target minerals: alunite, kaolinite, and montmorillonite. The top row of plots show the regions where the target mineral shows higher abundance and the bottom row depicts the actual selected pixels from the locations with higher presence of the target mineral.}
 \label{fig:ROI.png}
\end{figure}

The specific regions of interest (ROIs) within the AVIRIS image are comprised of pixels from alunite, kaolinite and montmorillonite locations. We used a graphical user interface, ``Hyperspectralpy" \cite{Basener_2022b} and the adaptive coherence estimator (ACE) detection algorithm to select the pixels within the ROIs. ``Hyperspectralpy" is a Python package that provides enhanced analysis and visualization of hyperspectral images. The ACE algorithm selects target mineral spectra from the spectral library and applies a whitening transformation to the image and library spectra. Following the whitening process, an ACE score is computed depicting the strength of the target mineral's presence within image. Figure \ref{fig:ROI.png} displays the selected ROIs via the red colored regions. The top row in Figure \ref{fig:ROI.png} are the plots of the ACE detection score for the target mineral. An increased presence of alunite, kaolinite and montmorillonite are indicated by the dark red. Consequently, we selected pixels with higher computed ACE scores (shown in the bottom row of Figure \ref{fig:ROI.png}).

We then inferred models from the selected mixed pixels. Each inferred model represents the best fit of the observed spectrum. Subsequently, we recorded the category of each mineral from every model. Mineral categories are generally based on the physical-chemical or atomic bonding structure of the substance. These characteristics have both distinctive as well as similar properties ultimately modifying their utility in material identification.  

\begin{figure*}[ht] 
\begin{subfigure}{.49\textwidth}
\includegraphics[width=\textwidth]{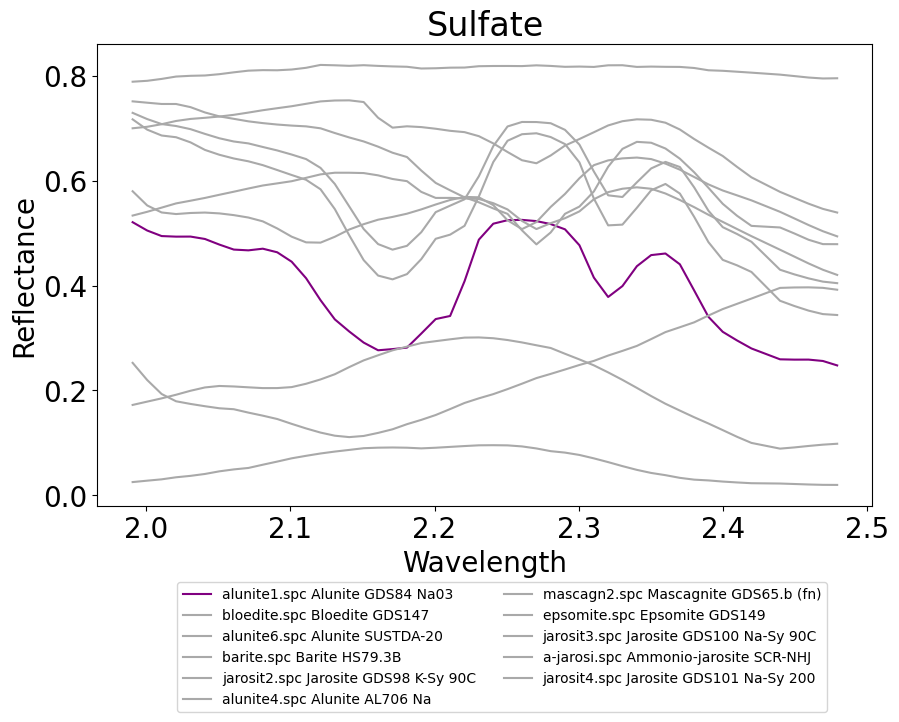}
\caption{Chemical Formula: $\text{SO}_4^{2-}$}
\label{fig:Sulfate}
\end{subfigure}
\begin{subfigure}{.49\textwidth}
\includegraphics[width=\textwidth]{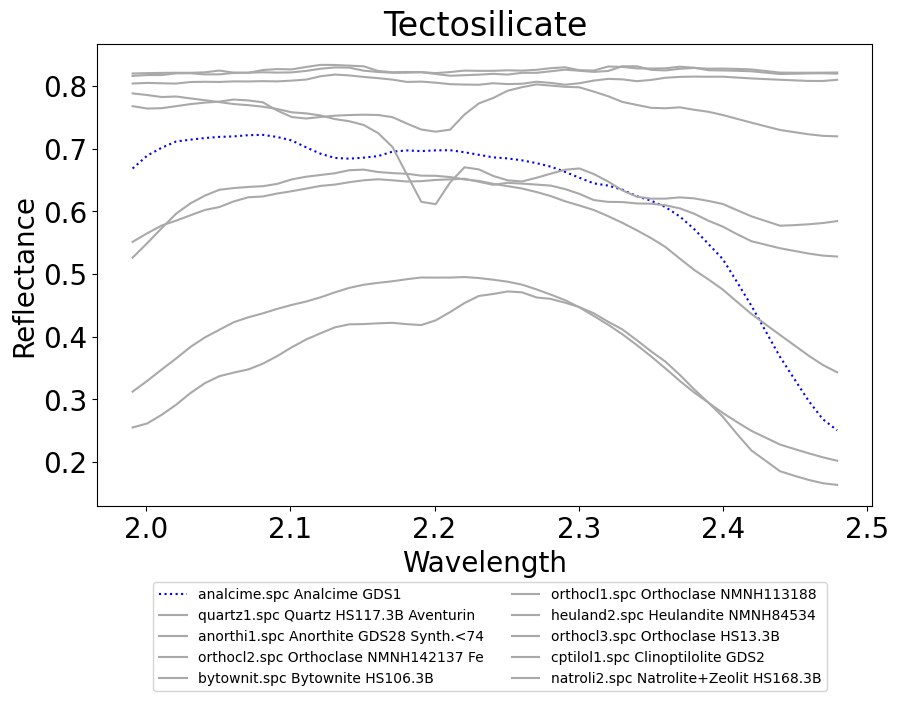}
\caption{Chemical Formula:  $\text{SiO}$}
\label{fig:tectosilicate}
\end{subfigure}
\begin{subfigure}{.49\textwidth}
\includegraphics[width=\textwidth]{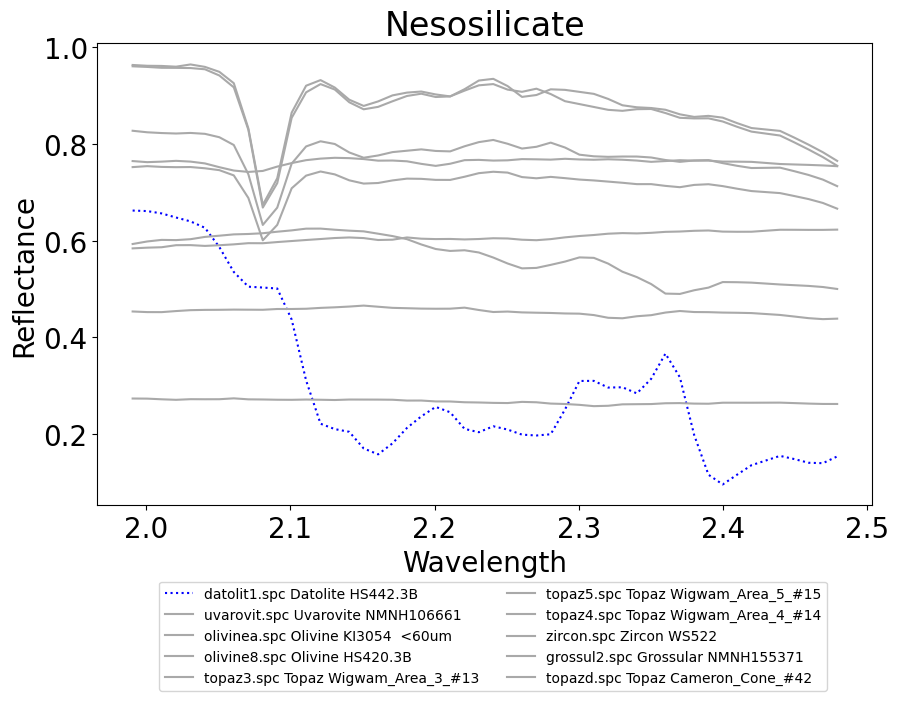}
\caption{Chemical Formula: $\text{SiO}_4$}
\label{fig:nesosilicate}
\end{subfigure}
\begin{subfigure}{.49\textwidth}
\includegraphics[width=\textwidth]{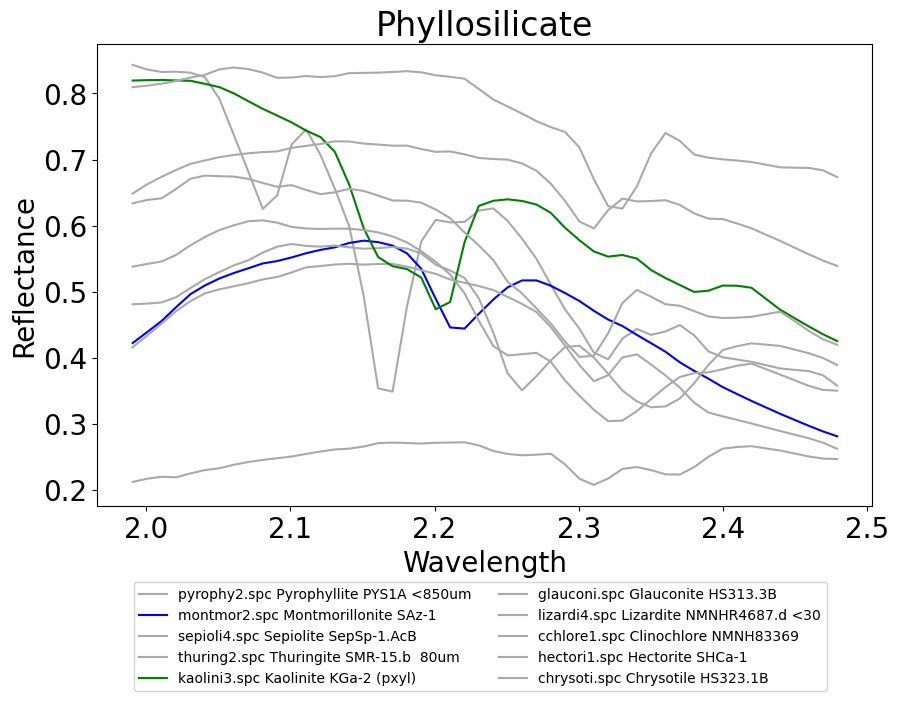}
\caption{Chemical Formula: $\text{Si}_2\text{O}_5$}
\label{fig:Phyllosilicate}
\end{subfigure}
\caption[example] { \label{fig:physical-chemical 1} 
Comparison of the mineral types sulfate, tectosilicate, and nesosilicate, and phyllosilicate. The purple spectrum (Figure \ref{fig:Sulfate}) represents alunite. The green and blue spectra (Figure \ref{fig:Phyllosilicate}) represent kaolinite and montmorillonite, respectively. The dotted blue spectra within the other plots (Figures \ref{fig:tectosilicate} and \ref{fig:nesosilicate}) are minerals that were often incorporated in the models. }
\end{figure*}

Figure \ref{fig:physical-chemical 1} displays a few of the most common mineral categories of the spectra incorporated in each of the models. Each plot depicts the spectral pattern of minerals from specific categories. The physical patterns are indicative of their chemical structure. Sulfates have a $\text{SO}_4^{2-}$ chemical formula because they are comprised of one sulfur and four oxygen atoms. Tectosilicates have a $\text{SiO}$ chemical formula which is indicative of their silicon and oxygen composition. Nesosilicates generally incorporate a silicon atom and four oxygen atoms, $\text{SiO}_4$. Phyllosilicates have a chemical formula, $\text{Si}_2\text{O}_5$, describing their composition of two silicon atoms and five oxygen atoms. 

\subsection{Statistical Analysis}

\subsubsection{LASSO Regression}
LASSO regression was incorporated in this study because of its sparse solutions and known computational advantages. It responds to the disadvantages of OLS because it minimizes overfitting through noise reduction. The penalty parameter controlling sparsity was selected using cross validation (CV) ranging from 0.001. to 0.1 with 0.001 step size and k=5 fold. Additionally, we used the Python packages  ``Lasso" and ``LassoCV" for implementation.

\subsubsection{Depth-first Search Strategy}
We implemented a DFS strategy because of its inherent feature selection capabilities. We developed the script for this technique and did not utilize a Python package. This approach can incur higher computational risk due to its iterative nature, but this allows for primary variable identification. 

\subsubsection{Mixed Integer Nonlinear Program}
The formulation of the MINLP within this paper has similarities to the models developed in Bourguignon et al. and Mhenni et al. with slight deviations \cite{bourguignon2015exact,mhenni2018spectral}. The following model is the mathematical program incorporated in this study.

\textbf{Sets}
\begin{align*}
    I &= \text{Set of all spectra in the library} \quad  i = [1,...,n] \in I \\    
    J &= \text{Set of all spectral bands} \quad  j = [1,...,m] \in J
\end{align*}

\textbf{Parameters}
\begin{align*}
    y_{j} &= \text{observed pixel reflectance for band \emph{j}}\\
    s_{ij} &= \text{reflected intensity for spectrum \emph{i} in band \emph{j}}\\
    M &= \text{number of spectral bands}\\
    N &= \text{number of library spectra}\\
     B_{i} &= \text{max abundance associated with spectrum \emph{i}}\\
     P &= \text{number of spectra included in the model}
\end{align*}

\textbf{Decision Variables}
\begin{align*}
    x_{i} &= \text{decision to assign spectrum \emph{i} to the model}\\
    a_{i} &= \text{abundance associated with spectrum \emph{i}}\\
\end{align*}

\textbf{Objective Function}
\begin{align*}
minimize \hspace{.1in} Z= \sqrt{\frac{1}{M}\sum_{j=1}^{M}{\bigg(y_{j}-\sum_{i=1}^{N}(a_{i}s_{ij}})\bigg)^{2}}
\end{align*}

\textbf{Constraints}
\begin{align*}
B_{i}x_{i} & \geq a_{i}\\
\sum_{i}^{n}x_{i} & \geq P
\end{align*}

\begin{align*}
\forall x_{i}\in \{0,1\}, \quad a_{i} \geq 0
\end{align*}

\subsection*{Objective Explained}
\begin{center}
Minimize the root mean squared error between the observed spectrum and the inferred spectrum (the spectra incorporated in the model and their associated abundances).
\end{center}

\subsection*{Constraints Explained}
\begin{enumerate}
\item Abundances can only be assigned to spectra included in the model and must adhere to the maximum material reflectance.

\item Model size constraint: The sum of the number of spectra assigned to the model must adhere to the model size parameter (P).

\item The decision to include a spectrum in the model ($s_{i}$) is binary. The abundance associated with a spectrum ($a_{i}$) must be positive.
\end{enumerate}

\subsubsection{HySUDeB}
In this novel approach, we decorrelated the spectral bands as part of the unmixing process. This band decorrelation is also known as whitening. The OLS assumption is that the residuals follow a multivariate normal distribution with the mean centered at the pixel spectrum and covariance proportional to the identity matrix. However, this assumption may not always be appropriate, thus whitening becomes beneficial because we assume the covariance matrix of the spectral bands is not originally constant. Within the whitening process, we computed the covariance of our observed image. From the covariance matrix we derived the eigenvalues and eigenvectors. The library spectra and the observed spectrum were whitened by subtracting the mean and applying a transformation using the covariance matrix's eigenvalues and eigenvectors \cite{basener2011automated}.  Once band decorrelation was complete, we applied LASSO regression to the whitened spectra.

\section{Results}
\begin{figure}
   \begin{center}
\includegraphics[width=.95\textwidth]{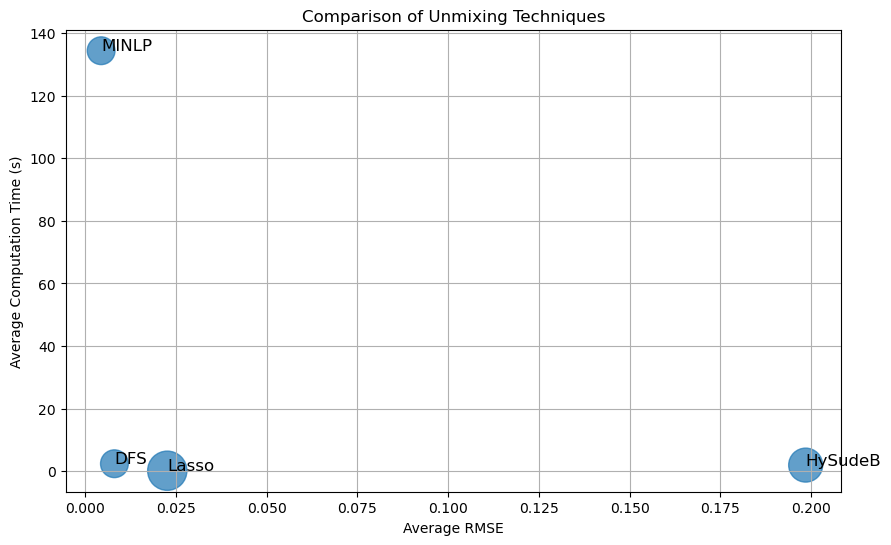}
	\end{center}
 \caption{Display of technique performance based on average error and computation time for the alunite ROI.}
 \label{fig:TechniqueComparison2.png}
\end{figure}

We compared the unmixing performance of each technique based on computation time, error, detection accuracy and mean average precision at k. Detection accuracy was based on the number of target spectra inferred from the modeling approaches -- we tallied the number of alunite, kaolinite, and montmorillonite detections achieved by each approach. Figure \ref{fig:TechniqueComparison2.png} displays each technique's performance for the alunite ROI in terms of computation time and root mean squared error (RMSE) -- square root of equation \ref{eq:MSE_equation}. Though Figure \ref{fig:TechniqueComparison2.png} is the result for the alunite ROI and the resulting plots for kaolinte and montmorillonite have different numeric values, the performance spread is the same for each of the ROIs in terms of this average RMSE vs average computation time plot. LASSO regression had the lowest average computation time. MINLP had the lowest average residual error, but far higher computation time than the other three techniques. DFS feature selection was not the highest performer in either area, but it shows to be the best overall for speed and minimal error. HySUDeB had similar computation time as LASSO regression, but because the units on the data for HySUDeB are different than the units for the other methods, the computed RMSE was much higher. 

The results for detection accuracy are shown in Table \ref{overall-detection}. For each of the ROIs, LASSO regression incorporated the the highest number of target minerals across each of its models. MINLP had the second highest detection percentages. When target variables were incorporated in the solutions, MINLP only included the target mineral once -- while other approaches incorporated variations of the target in many models. 

Table \ref{tab:Precision.png} displays the results in terms of detection precision.  Precision at k is an indication of importance placed on the target mineral. Minerals included in each model were ranked based on model size and assigned abundance. Specifically, precision at k measures the number of times the target mineral was one of the top ordered materials. This precision table shows that HySUDeB performed the best with the alunite ROI and MINLP preformed the best with the other two regions, kaolinite and montmorillonite.

\begin{table}[ht]
\centering
\begin{tabular}{lllll}
\multicolumn{5}{c}{Target Detection Percentage} \\ \hline
\multicolumn{1}{l|}{} & \multicolumn{1}{l|}{Lasso} & \multicolumn{1}{l|}{DFS} & \multicolumn{1}{l|}{MINLP} & HySudeB \\ \hline
\multicolumn{1}{l|}{Alunite} & \multicolumn{1}{l|}{1.0} & \multicolumn{1}{l|}{1.0} & \multicolumn{1}{l|}{1.0} & 0.95 \\ \hline
\multicolumn{1}{l|}{Kaolinite} & \multicolumn{1}{l|}{1.0} & \multicolumn{1}{l|}{0.65} & \multicolumn{1}{l|}{1.0} & 0.94 \\ \hline
\multicolumn{1}{l|}{Montmorillonite} & \multicolumn{1}{l|}{0.85} & \multicolumn{1}{l|}{0.56} & \multicolumn{1}{l|}{0.77} & 0.22 \\ \hline
\end{tabular}
\caption{\label{overall-detection}Detection percentages for target minerals (alunite, kaolinite, and montmorillonite) across four techniques: Lasso, DFS, MINLP, and HySudeB.}
\end{table}

\begin{table}[ht]
\centering
\begin{tabular}{lllll}
\multicolumn{5}{c}{Mean Average Precision at k} \\ \hline
\multicolumn{1}{l|}{} & \multicolumn{1}{l|}{Lasso} & \multicolumn{1}{l|}{DFS} & \multicolumn{1}{l|}{MINLP} & HySudeB \\ \hline
\multicolumn{1}{l|}{Alunite} & \multicolumn{1}{l|}{0.41} & \multicolumn{1}{l|}{0.58} & \multicolumn{1}{l|}{0.36} & 0.60 \\ \hline
\multicolumn{1}{l|}{Kaolinite} & \multicolumn{1}{l|}{0.39} & \multicolumn{1}{l|}{0.31} & \multicolumn{1}{l|}{0.54} & 0.46 \\ \hline
\multicolumn{1}{l|}{Montmorillonite} & \multicolumn{1}{l|}{0.14} & \multicolumn{1}{l|}{0.14} & \multicolumn{1}{l|}{0.18} & 0.03 \\ \hline
\end{tabular}
\caption{\label{tab:Precision.png}Average mean precision at k for minerals (alunite, kaolinite, and montmorillonite) across four techniques: Lasso, DFS, MINLP, and HySudeB.}
\end{table}

\begin{figure}
   \begin{center}
\includegraphics[width=.58\textwidth]{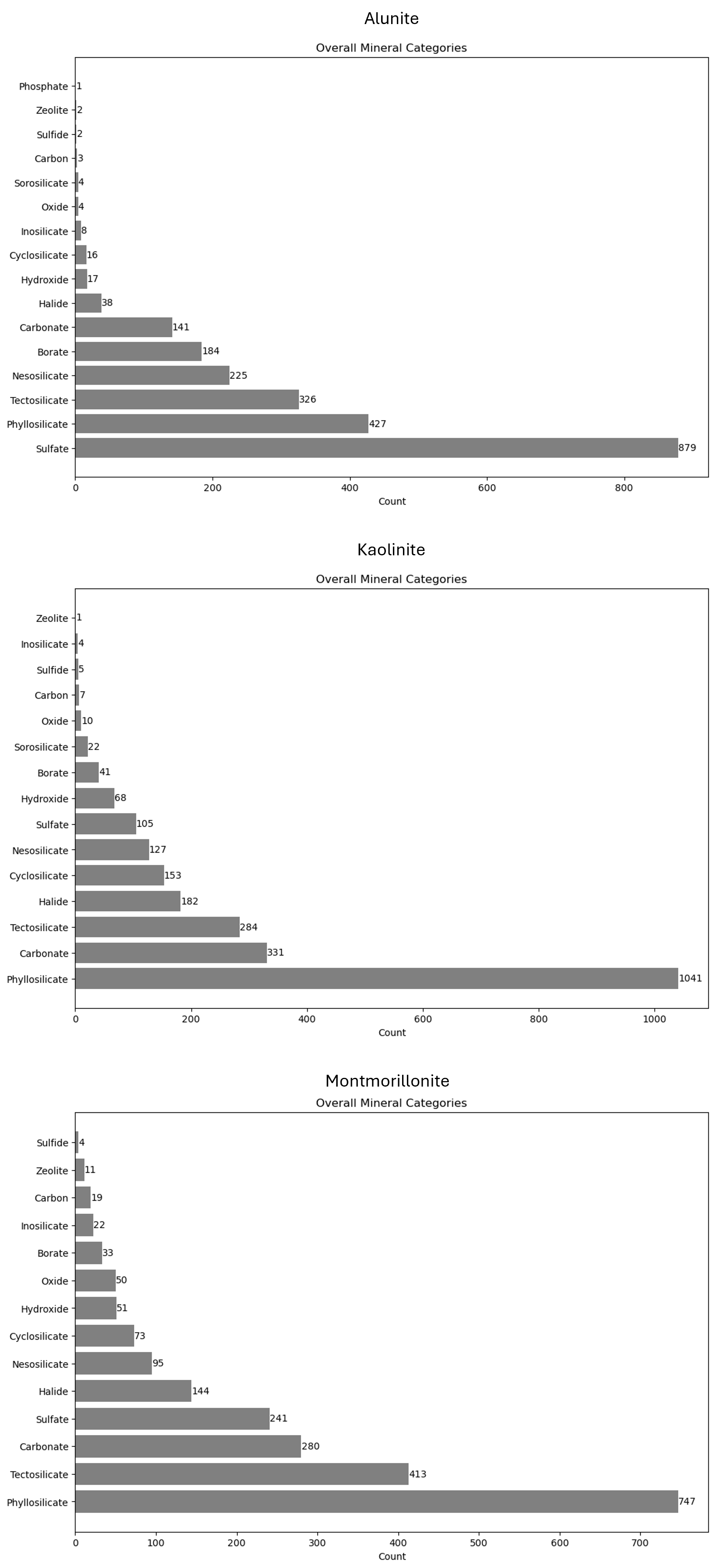}
	\end{center}
 \caption{Display of the count of the included minerals across all the implemented approaches.}
 \label{fig:CompHistograms.png}
\end{figure}

\begin{figure}
   \begin{center}
\includegraphics[width=\textwidth]{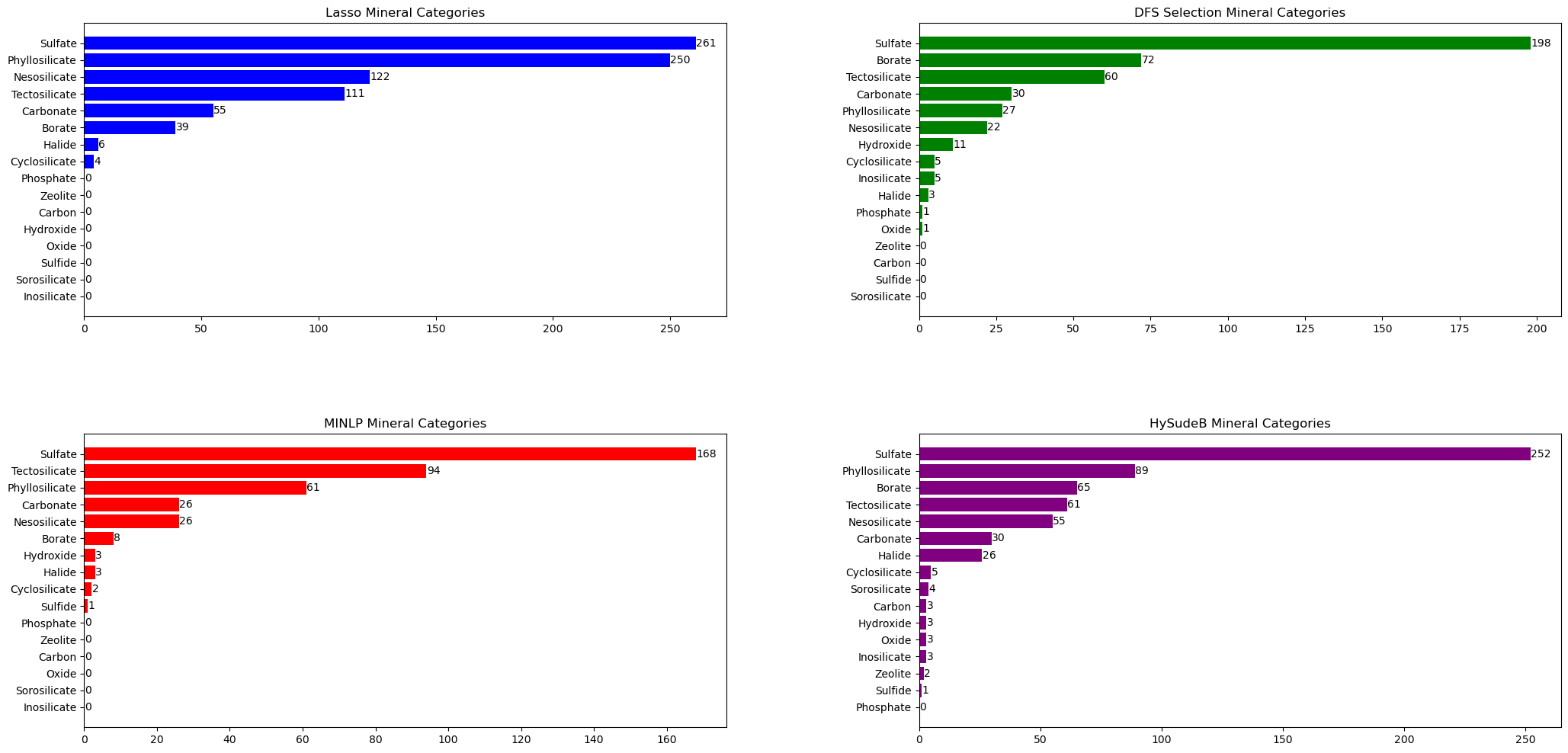}
	\end{center}
 \caption{Display of the count of the minerals from the library that were included in the models for the alunite ROI.}
 \label{fig:TechniqueHistograms.png}
\end{figure}

In addition to measuring performance, we identified the mineral categories incorporated in the each modeling approach. The most commonly included categories regardless of technique were sulfates, tectosilicates, nesosilicates, carbonates and phyllosilicates. Figure \ref{fig:CompHistograms.png} shows the count for all the included mineral categories across all three ROIs. The categories shown in Figure \ref{fig:CompHistograms.png} are not an exhaustive list of the mineral types found in the spectral library. Figure \ref{fig:TechniqueHistograms.png} shows a more extensive list of the primary library categories as well as the counts based on unmixing approach. LASSO regression showed to have the most difference in mineral inclusion but the lowest diversity in mineral category for each ROI. For example in Figure \ref{fig:TechniqueHistograms.png}, Hydroxide minerals were included in all the other approaches expect LASSO. DFS feature selection and HySUDeB had similar spreads of included mineral categories. Both approaches show a definitive gap between the highest counts and second highest counts. Even though, LASSO regression, included the highest amount of sulfates in the models for the alunite ROI (presumably because alunite is a sulfate), LASSO regression included nearly as many phyllosilicates. HySUDeB was a close second to LASSO regression in terms of icluding sulfates in the model for the alunite ROI, however unlike LASSO regression there is a distinct difference between the count for sulfates and phyllosilicates. HySUDeB shows this pattern of category inclusion with the kaolinite and montmorillonite ROIs as well.  

\section{Discussion}
This study compared four unmixing techniques using four aspects: computation time, error minimization, detection accuracy and mean average precision at k. Our results indicate that the employed techniques dominate in specific tasks and goals. No one method was universally superior but each are highly effective depending on the aim of the study. LASSO regression was the fastest unmixing technique. MINLP had the lowest minimal error and DFS feature selection achieved the best balance in every category. HySUDeB performed well in detecting the target mineral category and thus features associated with those categories. 

The goal of this study was primary mineral detection. LASSO regression performed the best due to cross validation. The penalty parameter allowed the model to become less sparse. This lowered sparsity ensured: 1) the target mineral was included in the model at a minimum 85\% of the time and 2) LASSO regression obtained the highest cumulative target mineral count. However, this also caused the model sizes to be larger and ultimately lowered the mean average precision at k. 

HySUDeB performed well in terms of feature detection of the target mineral -- as did DFS feature selection. For each of the ROIs, the chemical category of the target mineral was definitive because HySUDeB addresses the OLS homoscedasticity failures.  In this approach, we decorrelated spectral bands, effectively whitening the data. By transforming the spectral data into a space where the bands are uncorrelated, we stabilized the variance of the residuals and improved the reliability of our model estimates. 

The decorrelation process in HySUDeB enhanced the precision scoring of specific minerals, such as alunite. Band decorrelation ensured evenly distributed error, enabling more precise identification of mineral signatures in the presence of varying noise levels across spectral bands. Computation time was not affected with the implementation of whitening, but the error had different units and thus comparing RMSE was not meaningful. Future work could focus on refining techniques like HySUDeB to decrease the residual error while also maintaining the computation and detection capabilities.

\section*{Acknowledgments}
The authors would like to thank the School of Data Science at the University of Virginia for their support during this research process and providing access to publication catalogs. 

\printbibliography

\appendix 
\section{Referenced code}
All the referenced and supporting code will be stored within the corresponding GitHub repository (\url{https://github.com/bakerjf1993/Hyperspectral-Unmixing/tree/main/2024%20Journal%20of%20Remote%20Sensing}).

\end{document}